\documentstyle[aps,twocolumn]{revtex} 

\newcommand{\lsim}{ \raisebox{-.5ex}{\mbox{$\,\stackrel{<}{\sim}$\,}} }
\begin{document}  
\draft
\title{How long does the quantum chaos last?}                               
\author{Giulio Casati}  
\address{Universit\`a di Milano -- Sede di Como,\\ Via Castelnuovo, 7
 -- 22100 Como, Italy
}
\author{B.V. Chirikov and O.V. Zhirov} 
\address{Budker Institute of Nuclear Physics\\
630090 Novosibirsk, Russia
}
\maketitle     
\begin{abstract}  
The main purpose of this Comment is to point out that besides the short time 
scale of quantum chaos, confirmed once more in recent paper by Alicki et al,
there is generally another time scale $t_R$, which is much longer and on which 
a partial quantum--classical correspondence persists. Namely, the quantum 
diffusion closely follows the classical one even though the former is 
dynamically stable. The absence of the long scale $t_R$ in model studied by 
Alicki et al. is a result of a special choice for one  model's parameters value.
\end{abstract}


\par                                                                            
\vspace{0.5cm}

In a recent paper[1] Alicki, Makowiec and Miklaszewski presented 
one more confirmation for the existence of the short (logarithmic)
time scale in quantum chaos using a simple model of kicked quantum top
and a finite--time analogue of the classical dynamical Kolmogorov - Sinai
(KS) entropy.
This random time scale $t_r\sim \ln{\hbar^{-1}}$, on which the quantum motion
is fully similar to the classical one including the exponential instability,
had been discovered in Ref.[2] and was subsequently confirmed and further
studied in many papers[3--5] (see also Ref.[6]).

The main purpose of this Comment is to point out that besides this 
short time scale there is generally another one, $t_R$, which is much longer
($\ln{t_R}\sim \ln{\hbar^{-1}}$) and on which a partial quantum--classical
correspondence persists, namely, the quantum diffusion closely
follows the classical one even though the former is dynamically stable
[7] (see also Refs.[3,5]). The absence of the long scale $t_R$ in model [1]
is a result of a special choice for one  model's parameters value.

Generally, the quantum top is described [8] by the unitary operator
(per kick, $\hbar =1$):
$$
   U(p,\,k,\,j)\,=\,{\rm e}^{-ikJ^2_z/2j}\,{\rm e}^{-ipJ_y} \eqno (1)
$$
which depends on two classical parameters, $k$ and $p$, and one quantum
parameter $j\gg 1$ (in quasiclassics). In Ref.[1] the value  $p=\pi /2$
was chosen following Ref.[8] in which the only reason for such a choice was
merely to simplify the quantum map. This particular choice leads to a
 nongeneric,
fast ('ballistic') relaxation to the ergodic steady state. According to
data in Fig.2 [1] the relaxation time $t_{er}(p)\approx 1.5$ 
iterations only,
in this case. Moreover, some relaxation occurs even for almost regular motion
($k=1$, see Fig.1 in Ref.[1]). 

On the contrary, if $p\ll 1$ the relaxation becomes diffusive and 
relatively slow, and only for chaotic motion, of course, namely
when parameter $K=pk>1$. 
In the simplest case $|J_z|\ll j$ the diffusion rate
in $J_z$ is [8,9]
$$
   D\,\approx\,\frac{1}{2}\,(pj)^2\,C(K)\,\sim\,(pj)^2 \eqno (2)
$$
where $C(K)\sim 1$ accounts for dynamical correlations. Hence, the 
relaxation time (in number of kicks) is
$
   t_{er}(p)\,\sim\,j^2/D\,\sim\,p^{-2}\,\gg\,1 
$.
During the relaxation process the quantum entropy keeps growing until
 it reaches the maximal value $H_{er}$ for the ergodic state:
$$
   H(t)\,\equiv\,-\,\sum\,f(J_z,\,t)\,\ln{f(J_z,\,t)}\,\to\,H_{er}\,=\,
   \ln{(2j\,+\,1)}    \eqno (3)
$$
where $J_z$ are integers, and $f(J_z)=|\psi (J_z,\,t)|^2$ is the
distribution function. 
An extra factor 2 in Ref.[1] (see Eq.(8) and Fig.1) for $H_{er}$ is not
completely clear but this is not the main point of our
Comment, and will not be discussed here.

For an initially narrow Gaussian distribution the entropy in the diffusion
 regime is
$
   H_D(t)\,\approx\,\ln{(2\pi{\rm e}Dt)}/2 
$
assuming $J_z\gg 1$ and $D(J_z)\approx const$ [9]. This entropy growth
is much slower than that on the random time scale $t<t_r$ ($H_r(t)=t\cdot h_r$), 
and the corresponding KS entropy 
vanishes [10] as it should be for a quantum motion with discrete spectrum.   

Notice that in the classical limit the entropy would grow indefinitely
with constant rate
$
   h_r\,=\,\Lambda\,\approx\,\ln{(K/2)} 
$
due to continuity of variable $J_z$  as explained in the beginning
of Ref.[1] (see also Ref.[11]). In the quantum case the classical
instability $h_r$ is restricted to the short time scale $t_r$ which can be
approximately found from the equation: $t\cdot h_r(t_r)=H_D(t_r)$. This
gives a new asymptotic ($j\to\infty$) estimate
$
   t_r\,\approx\,\ln{(pj)}/\ln{(pk)}
$
in agreement with previous results [2--5].

The final steady state is ergodic with entropy (3) only under the additional 
condition [3,5,9]
$ t_{er}\,\ll\,t_H\,=\,(2j\,+\,1)/2\pi\,=\,\exp{(H_{er})}/2\pi$ 
or $jp^2\,\gg\,1 $
where $t_H$ is the mean quasienergy level density also called the
Heisenberg time. In the opposite case ($jp^2\lsim 1$) the quantum diffusion
is restricted to the relaxation (diffusion) time scale[3,5,9]
$$
   t_R\,\sim\,D\,\sim\,(pj)^2\,\lsim\,t_H \eqno (4)
$$
Hence, the quantum steady state is essentially nonergodic due to localization
of quantum diffusion. Assuming approximately exponential
localization with a characteristic length $l\approx D$ the final steady
state entropy in this case is
$
  H_l\,\approx\,1\,+\,\ln{D}\,\to\,2\,\ln{(pj)}\,\lsim\,H_{er} 
$

The diffusive time scale $t_R$ (4), which is the main point of our Comment,
is always much longer as compared to the instability scale $t_r$.
Only for $t\gg t_R$ the motion is completely dominated by the quantum effects.

\end{document}